\documentclass[prb,
reprint,
floatfix,
superscriptaddress,
]{revtex4-1}

\usepackage{graphicx,amsmath,amssymb,bm}
\newcommand{\sxc}{_{\mathrm{xc}}}
\newcommand{\fxc}{f_{\mathrm{xc}}}
\newcommand{\rs}{r_{\mathrm{s}}}
\newcommand{\kf}{k_{\mathrm{F}}}
\newcommand{\br}{\bm{r}}
\newcommand{\kb}{\overline{k}}
\newcommand{\bg}{\bm{G}}
\newcommand{\bq}{\bm{q}}
\newcommand{\re}{\mathrm{Re}\,}
\newcommand{\im}{\mathrm{Im}\,}

\usepackage[dvipsnames]{xcolor}

\begin{document}

\title{Progress towards understanding ultranonlocality through the wavevector and frequency dependence of approximate exchange-correlation kernels}

\author{Niraj K. Nepal}
\affiliation{Department of Physics, Temple University, Philadelphia, PA 19122}
\author{Aaron D. Kaplan}
\affiliation{Department of Physics, Temple University, Philadelphia, PA 19122}
\author{J. M. Pitarke}
\affiliation{CIC nanoGUNE BRTA and DIPC, E-20018 Donostia, Basque Country, Spain}
\affiliation{Materia Kondentsatuaren Fisika Saila and Centro Fisica Materiales CSIC-UPV/EHU, E-48080 Bilbao, Basque Country, Spain}
\author{Adrienn Ruzsinszky}
\email{Author to whom correspondence should be addressed: tuf27796@temple.edu}
\affiliation{Department of Physics, Temple University, Philadelphia, PA 19122}

\date{\today}

\begin{abstract}
  In the framework of time-dependent density functional theory (TDDFT), the exact exchange-correlation (xc) kernel $\fxc(n,q,\omega)$ determines the ground-state energy, excited-state energies, lifetimes, and the time-dependent linear density response of any many-electron system. The recently developed MCP07 xc kernel $\fxc(n,q,\omega)$ of A. Ruzsinszky \textit{et al.} [Phys. Rev. B {\bf 101}, 245135 (2020)] yields excellent uniform electron gas (UEG) ground-state energies and plausible plasmon lifetimes. As MCP07 is constructed to describe $\fxc$ of the UEG, it cannot capture optical properties of real materials. To verify this claim, we follow Nazarov {\it et al.} [Phys. Rev. Lett. {\bf 102}, 113001 (2009)] to construct the long-range, dynamic xc kernel, $\lim_{q\to 0}\fxc(n,q,\omega) = -\alpha(\omega)e^2/q^2$, of a weakly inhomogeneous electron gas, using MCP07 and other common xc kernels. The strong wavevector and frequency dependence of the ``ultranonlocality'' coefficient $\alpha(\omega)$ is demonstrated for a variety of simple metals and semiconductors. We examine how imposing exact constraints on an approximate kernel shapes $\alpha(\omega)$. Comparisons to kernels derived from correlated-wavefunction calculations are drawn.
\end{abstract}

\maketitle

\section{Introduction}

By virtue of the Runge-Gross theorem \cite{runge1984}, time-dependent density functional theory (TDDFT) is an extension of ground-state density functional theory (DFT) \cite{gross1985,ullrich2012,ullrich2014,maitra2016}. TDDFT is a more computationally feasible approach to compute excitation energies, compared to approaches based on many-body techniques \cite{rohlfing2000,hanke1979}.

The response of a many-electron system to a dynamic, external potential is characterized by a change in its charge density. If the amplitude of the external potential is small, then it exerts only a weak perturbation, and the density response can be taken to be linear. Through the linear density-density response function, linear response TDDFT (LR-TDDFT) \cite{petersilka1996} can predict the transition frequencies to electronic excited states, among other properties.

The central equation of LR-TDDFT is a Dyson-like equation linking the true interacting density-response function of an arbitrary many-electron system, $\chi(\br,\br';\omega)$, to its non-interacting Kohn-Sham (KS) counterpart $\chi_0(\br,\br';\omega)$:
\begin{align}
  \chi(\br,\br';\,&\omega) = \chi_0(\br,\br';\omega) + \int d^3 r'' \int d^3 r''' \chi_0(\br,\br'';\omega) \nonumber \\
  & \times \left[v(\br'',\br''') + \fxc(\br'',\br''';\omega) \right]  \chi(\br''',\br';\omega), \label{eq:chi_dyson}
\end{align}
where $v(\br,\br')$ represents the Coulomb interaction, and $\fxc(\br,\br';\omega)$ is the exchange-correlation (xc) kernel. The TDDFT excitation energies are found as poles of $\chi(\br,\br';\omega)$. With $\fxc=0$, $\chi$ reduces to the random-phase approximation (RPA) \cite{nozieres1958}.

Even if TDDFT is computationally more efficient than wavefunction theories, its accuracy is restricted by the limitations of commonly used xc kernels \cite{patrick2015,bates2016,constantin2007,terentjev2018}. The most widely used adiabatic local-density approximation (ALDA), for example, misses excitonic effects completely \cite{gross1996}. Recently developed model xc kernels describe the optical absorption spectra of small and medium band-gap semiconductors well \cite{patrick2015,terentjev2018}. Non-empirical xc kernels, which are not fitted to experimental or correlated-wavefunction data, are typically insufficient to describe excited-state properties of semiconductors and insulators, especially large-gap materials. However, some modern kernels predict accurate properties of excitons in real materials \cite{trevisanutto2013,rigamonti2015,berger2015,yang2015}. In order to achieve this goal, the wavevector and frequency dependence of the xc kernel $\fxc$ needs to be investigated further \cite{bruneval2006,botti2004,panholzer2018}.

Recently, we proposed a parametrized $\fxc$ for a uniform electron gas (UEG) based upon satisfaction of exact constraints \cite{ruzsinszky2020}. This model, called MCP07, depends upon both the wavevector $q$ and frequency $\omega$. In the static ($\omega=0$) limit, we modified the model of Constantin and Pitarke \cite{constantin2007} as follows:
\begin{equation}
  \fxc^{\text{MCP07}}(q,0) = \frac{4\pi}{q^2}B[e^{-k q^2}(1 + E q^4)-1] - \frac{4\pi}{\kf^2}\frac{C}{1 + (k q^2)^{-2}}. \label{eq:mcp07_static}
\end{equation}
For definitions of the $\rs$-dependent functions $k$, $B$, $C$, and $E$, refer to Ref. \cite{ruzsinszky2020}. $\rs=(4\pi n/3)^{-1/3}$ is the radius of a sphere containing, on average, one electron in a UEG of density $n$. $\kf = (3\pi^2n)^{1/3}$ is the Fermi wavevector. In the long-wavelength ($q\to 0$) limit, Eq. (\ref{eq:mcp07_static}) yields the ALDA kernel, for which we use the Perdew-Zunger parametrization of the UEG correlation energy \cite{perdew1981}.

At finite frequencies, we use the long-wavelength model $ \fxc (0,\omega)$ of Gross, Kohn, and Iwamoto (GKI) \cite{gross1985,iwamoto1987} (analytically continued to complex frequencies when needed). This expression for $\fxc(0,\omega)$ is then combined with the static limit of Eq. (\ref{eq:mcp07_static}) to obtain \cite{ruzsinszky2020}
\begin{equation}
  \fxc^{\text{MCP07}}(q,\omega) = \left\{ 1 + e^{-\kb q^2}\left[\frac{\fxc(0,\omega)}{\fxc(0,0)} - 1\right] \right\}\fxc^{\text{MCP07}}(q,0). \label{eq:mcp07_fxc}
\end{equation}
In MCP07, one takes $\kb=k$, where $k$ is the same function appearing in Eq. (\ref{eq:mcp07_static}). Here, we also discuss the effect of setting $\kb=0$. At long wavelengths ($q\to 0$), Eq. (\ref{eq:mcp07_fxc}) yields the {\it local} GKI $\fxc(0,\omega)$. At $\omega=0$, Eq. (\ref{eq:mcp07_fxc}) yields the {\it non-local} $\fxc^{\text{MCP07}}(q,0)$ of Eq. (\ref{eq:mcp07_static}).

The full MCP07 kernel of Eq. (\ref{eq:mcp07_fxc}) provides a highly-accurate description of ground-state correlation energies (up to $\rs=10$, see Figs. \ref{fig:ueg_eps_c} and \ref{fig:ueg_eps_c_errs}) and quasiparticle properties of the UEG. In particular, MCP07 predicts a finite plasmon lifetime that first decreases from infinity, and then increases as $q$ grows from 0 towards the electron-hole continuum. MCP07 also yields a static charge-density wave at $\rs \gtrsim 69$ that can be associated with a softening of the plasmon mode \cite{perdew1980,perdew2021}. The exchange-only version of the static MCP07 xc kernel of Eq. (\ref{eq:mcp07_static}) confirms Overhauser's prediction \cite{overhauser1968} that correlation is essential for the creation of a charge-density wave.

As MCP07 approximates $\fxc$ of the UEG, it cannot describe the $1/q^2$ long-wavelength behavior of the exact $\fxc$ for non-uniform many-electron systems, called ``ultranonlocality'' \cite{botti2004}. Indeed, in reciprocal space, the kernel is a matrix characterized by reciprocal wavevectors, whose spatial decay manifests in possible leading $q$-independent terms, called crystal local-field effects. The head and wings of adiabatic kernels, derived from semilocal density functional approximations, are independent of $q$, and thus are incorrectly non-divergent for $q \to 0$, as in the case of the UEG. The so-called ``bootstrap'' idea represents a very unique and effective route to account for ultranonlocality \cite{sharma2011}.

This work builds upon that of Nazarov {\it et al.} \cite{nazarov2009} by constructing, in the optical limit, the dynamic xc kernel of a weakly inhomogeneous electron gas using the MCP07 xc kernel (among other UEG-based kernels) as input. As MCP07 simultaneously describes the wavevector and frequency dependence of the xc kernel, our approach serves as a basis for further investigations of the optical absorption of non-uniform systems and the real wavevector and frequency dependence of xc kernels.

\section{What is known about the optical limit of the xc kernel? \label{sec:optical_limit}}

Giuliani and Vignale \cite{giuliani2005} provided a detailed discussion of exchange and correlation in uniform and non-uniform many-electron systems. They discussed, in particular, the important difference between the short-range kernel of the uniform electron gas, where $\fxc(q,\omega)$ tends to a finite constant as $q \to 0$, and the ultranonlocal kernel of non-uniform systems, where $\fxc$ is known to diverge in the long-wavelength limit as $1/q^2$.

Concurrent works have generalized kernel development from density to current-density functionals \cite{vignale1996}. Nazarov \textit{et al.} \cite{nazarov2007} derived a general method for constructing a scalar TDDFT xc kernel using the tensorial kernel and KS current-density response function of time-dependent current-density functional theory (TDCDFT) \cite{vignale1996,vignale1997}. This method proved to be particularly useful, as a local approximation to the xc kernel of TDCDFT results in a nonlocal approximation to the xc kernel of TDDFT. The resultant approximation is free of the contradictions that plague the standard local density approximation to TDDFT. This method also allowed the construction of the frequency-dependent xc kernel of a weakly inhomogeneous electron gas in the optical limit \cite{nazarov2009}:
\begin{equation}
 \lim_{q\to 0} \fxc(\bq,\bq,\omega) = -\frac{e^2\alpha(\omega)}{q^2}, \label{eq:fxc_alpha}
\end{equation}
where
\begin{equation}
  \alpha(\omega) =  -\frac{1}{e^2 \overline{n}^2_0 } \sum_{\bg \neq 0} (\bg \cdot \hat{\bq})^2[\fxc^{\text{HL}}(G,\omega) - \fxc^{\text{HL}}(G,0)]|n(\bg)|^2. \label{eq:alpha_sum}
\end{equation}
Here, $\hat{\bq}=\bq/|\bq|$, $n(\bg)$ is the Fourier transform of the electron density evaluated at the reciprocal lattice vector $\bg$, and $\overline{n}_0=n(\bg=\bm{0})$ is the average density. $\fxc^{\text{HL}}$ represents the longitudinal component of the tensor xc kernel of a uniform (homogeneous) electron gas, which coincides with its scalar counterpart; this uniform-gas xc kernel is evaluated at the average electron density $\overline{n}_0$. By averaging over all $\hat{\bq}$ directions, $(\bg \cdot \hat{\bq})^2$ is replaced by $G^2/3$. In the uniform limit, $\alpha(\omega)\to 0$. A negative sign is introduced in Eq. (\ref{eq:fxc_alpha}), as in Refs. \cite{botti2004, botti2005}, in such a way that a positive $\alpha(\omega)$ could cancel the divergent Coulomb interaction $4\pi/q^2$ as $q\rightarrow 0$.

In Ref. \cite{nazarov2009}, the xc kernel $\fxc^{\text{HL}}(q,\omega)$ entering Eq. (\ref{eq:alpha_sum}) was approximated as $\fxc^{\text{HL}}(q,\omega)\approx\fxc^{\text{HL}}(0,\omega) $, taking the latter from Ref. \cite{qian2002}. Here, we go a step further by using the non-local MCP07, which represents a reliable wavevector- and frequency-dependent uniform-gas xc kernel. From this dependence on the wavevector $\bg$, we expect that $\fxc(\bq,\bq,\omega)$ can cover new spectral features that are simply absent when $\fxc^{\text{HL}}(0,\omega)$ is used in Eq. (\ref{eq:alpha_sum}). At large wavevectors $q^2 \gg 1/\kb$, the frequency dependence of the MCP07 xc kernel is damped out significantly. In the limit $q\to \infty$, the MCP07 kernel approaches its static limit, therefore we expect a significant contribution to Eq. (\ref{eq:alpha_sum}) for small values of $\bg$ only. More details of the frequency dependence can be found in Ref. \cite{ruzsinszky2020}. On the other hand, our calculations indicate that the wavevector dependence of the uniform-gas xc kernel -- neglected in Ref. \cite{nazarov2009} -- can largely affect ultranonlocality.

Our calculation of the frequency-dependent coefficient $\alpha(\omega)$ of Eq. (\ref{eq:alpha_sum}) hinges upon the evaluation of: (i) the Fourier coefficients $n(\bg)$ and (ii) the MCP07 xc kernel of Eq. (\ref{eq:mcp07_fxc}), which is based on exact constraints.

Consider a UEG of density $\overline{n}_0 = \kf^3/(3\pi^2)$ perturbed by a weak periodic external pseudopotential
\begin{equation}
  W(\br) = \sum_{\bg} W(\bg) e^{i\bg \cdot \br}
\end{equation}
representing the actual crystal lattice. For  $W(\bg) = (\overline{n}_0/z)w(\bg)$, we apply the evanescent core pseudopotential $w$ of an ion of valence $z$ described by Eq. (2.11) of Ref. \cite{fiolhais1995}. The pseudopotential is designed to have a finite value at $r=0$, with vanishing first and third derivatives. This analytic behavior leads to a quick convergence of its Fourier transform in the limit $G \to \infty$.

Linear response tells us how to find $n(\bg)$:
\begin{equation}
  n(\bg) = \chi(\bg)W(\bg) = \frac{\chi_0(\bg)}{\epsilon(\bg)}W(\bg),
\end{equation}
with $\chi_0(\bg) = -\kf/\pi^2 F(G/(2\kf))$. $F$ is the static response function of the noninteracting UEG, known as the Lindhard function \cite{lindhard1954}:
\begin{equation}
  F(y) = \frac{1}{2} + \frac{1 - y^2}{4 y}\ln\left|\frac{1 + y}{1 - y} \right|.
\end{equation}
The screening in the UEG is represented by the dielectric function:
\begin{equation}
  \epsilon(\bg) = 1 - \left[\frac{4\pi}{\bg^2} + \fxc(\bg) \right]\chi_0(\bg), \qquad |\bg| > 0,
\end{equation}
where $\fxc(G)$ is the static xc kernel of the UEG.

The xc kernels $\fxc^\text{HL}(G,0)$ and $\fxc^\text{HL}(G,\omega)$ entering Eq. (\ref{eq:alpha_sum}) are taken from Eqs. (\ref{eq:mcp07_static}) and (\ref{eq:mcp07_fxc}), respectively.

\section{Results at the optical limit from local (LDA) and non-local (MCP07) dynamic kernels}

\begin{figure}
  \includegraphics[width=\columnwidth]{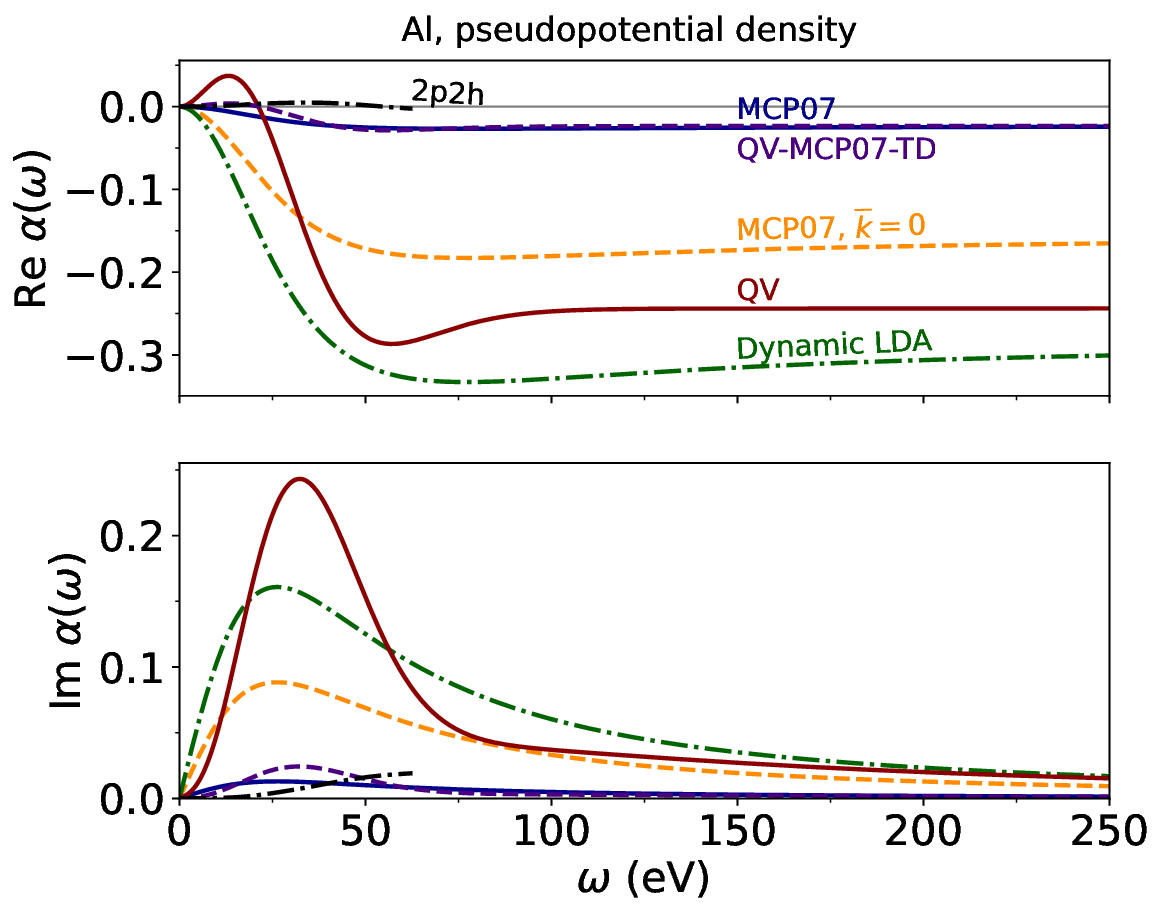}
  \caption{$\alpha(\omega)$ as a measure of ultranonlocality in fcc Al crystal. The upper (lower) figure shows the real (imaginary) part of the complex $\alpha(\omega)$ for the GKI dynamic LDA (green, dot-dashed), MCP07 with $\kb=0$
  (orange, dashed), the full MCP07 \cite{ruzsinszky2020} (blue, solid), frequency-dependent Qian and Vignale \cite{qian2002} (QV, red solid), QV-MCP07-TD (purple, dashed), and 2p2h \cite{panholzer2018} (black, dot-dashed) xc kernels. We are using $\fxc^{\text{HL}}(q,\omega)$, as shown in the calculation of $\alpha$. The labels of the curves indicate which $\fxc^{\text{HL}}(q,\omega)$ was used as input to Eq. (\ref{eq:alpha_sum}).
 \label{fig:al_alpha}}
\end{figure}

Figure \ref{fig:al_alpha} displays the coefficient $\alpha(\omega)$ of face-centered cubic (fcc) aluminum determined from either (i) Eq. (\ref{eq:alpha_sum}) with the fully non-local MCP07 $\fxc$ of Eq. (\ref{eq:mcp07_fxc}) as input, or (ii) the local density approximation (LDA) version of Eq. (\ref{eq:alpha_sum}):
\begin{equation}
  \alpha^{\text{DLDA}}(\omega) = -\frac{1}{3 e^2 \overline{n}^2_0 }\sum_{\bg \neq 0} \bg^2[\fxc^{\text{HL}}(0,\omega) - \fxc^{\text{HL}}(0,0)]|n(\bg)|^2.
  \label{eq:alpha_lda}
\end{equation}
$\fxc^{\text{HL}}(0,\omega)$ is taken to be either the GKI dynamic LDA xc kernel [the long-wavelength limit of Eq. (\ref{eq:mcp07_fxc})] or the Qian-Vignale (QV) dynamic LDA xc kernel of Ref. \cite{qian2002}. Note that the GKI and QV dynamic LDAs tend to distinct static limits; this is discussed further in Appendix B. Hereafter, we will use ``the dynamic LDA'' to refer to the GKI expressions. When using the fully non-local MCP07 xc kernel of Eq. (\ref{eq:mcp07_fxc}), the parameter $\kb$ is taken to be either equal to $k$, or equal to zero. We also introduce a hybrid kernel, which replaces the GKI frequency dependence in MCP07 with the QV model. As seen in Appendix B, ensuring that $\fxc(0,0)$ yields the ALDA yields more realistic correlation energies in the metallic range. This new kernel is called QV-MCP07-TD. All sums over $\bg$ used a sufficiently large cutoff of
$\bg_{\mathrm{c}}^2/2 < 800$ eV. We also present results for a correlated-wavefunction-derived kernel, as discussed after Eq. (\ref{eq:fxc_lrc}).

Our calculations clearly indicate that the coefficient $\alpha(\omega)$ is particularly sensitive to the wavevector dependence of the xc kernel, as seen by comparing MCP07 to its counterpart with the damping factor $\kb$ of Eq. (\ref{eq:mcp07_fxc}) set to zero. Physically, setting $\kb=0$ strengthens the frequency dependence of $\alpha(\omega)$ significantly over MCP07 (where $\kb=k$). Indeed, $|\alpha(\omega)|$ is reduced by more than 85\% when $\kb$ is increased from zero to its full MCP07 strength, $\kb=k$. These are consistent behaviors within the visible range of frequencies around 3 eV.

\begin{figure}
  \includegraphics[width=\columnwidth]{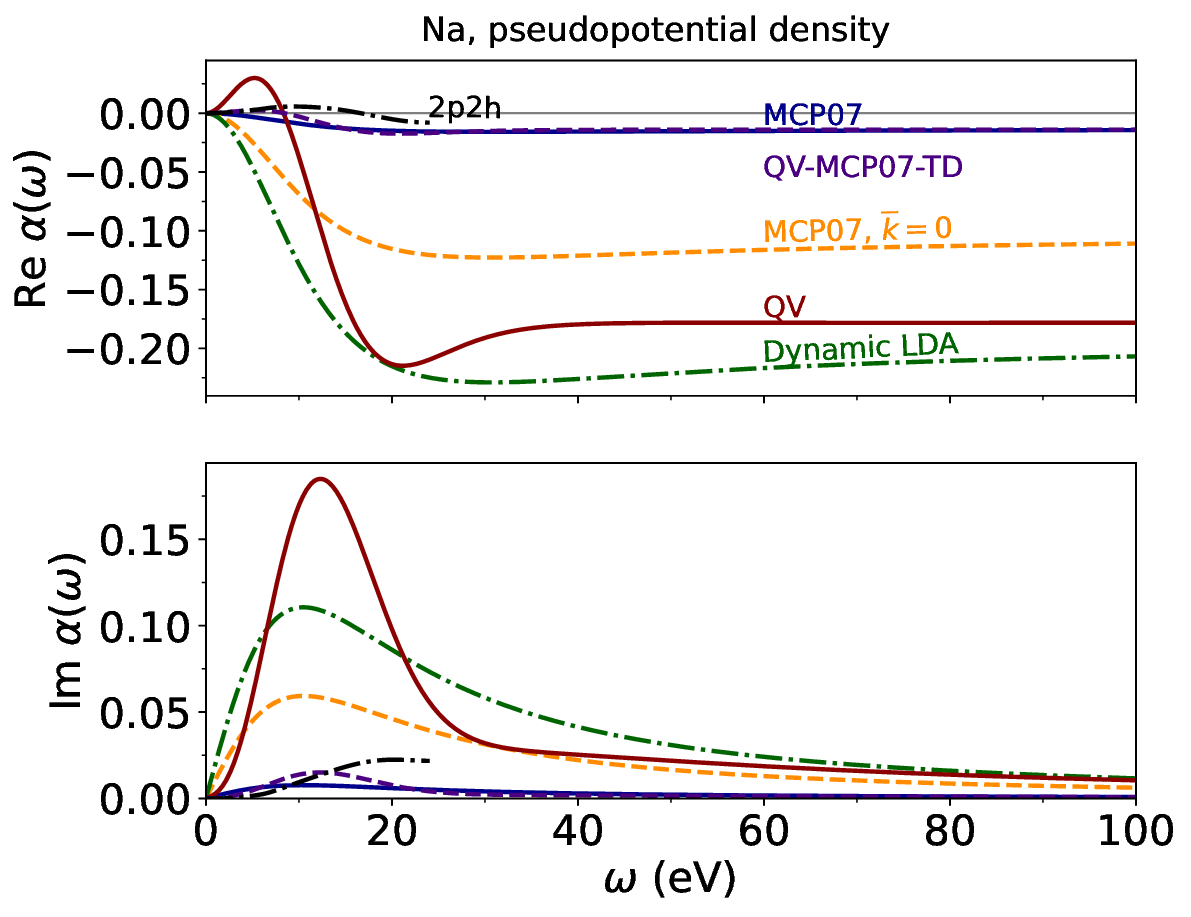}
  \caption{ $\alpha(\omega)$ as a measure of ultranonlocality in bcc Na crystal. The upper (lower) figure shows the real (imaginary) part of the complex $\alpha(\omega)$ for the GKI dynamic LDA (green dot-dashed), MCP07 with $\kb=0$ (orange dashed), the full MCP07 (blue solid) \cite{ruzsinszky2020}, frequency-dependent Qian and Vignale \cite{qian2002} (QV, red solid), QV-MCP07-TD (purple, dashed), and 2p2h \cite{panholzer2018} (black, dot-dashed) xc kernels.
 \label{fig:na_alpha}}
\end{figure}

The same physics is reported in Fig. \ref{fig:na_alpha} for body-centered cubic (bcc) Na. The coefficient $\alpha(\omega)$ is again particularly sensitive to whether the wavevector dependence of the xc kernel is considered. Note that our values shown in Figs. \ref{fig:al_alpha} and \ref{fig:na_alpha} for $\re \alpha(\omega)$ at optical frequencies differ considerably from the metallic limit ($\alpha=-0.213)$ of Eq. (4) of Ref. \cite{botti2004}, which was fitted to semiconductor data.

\begin{figure}
  \includegraphics[width=\columnwidth]{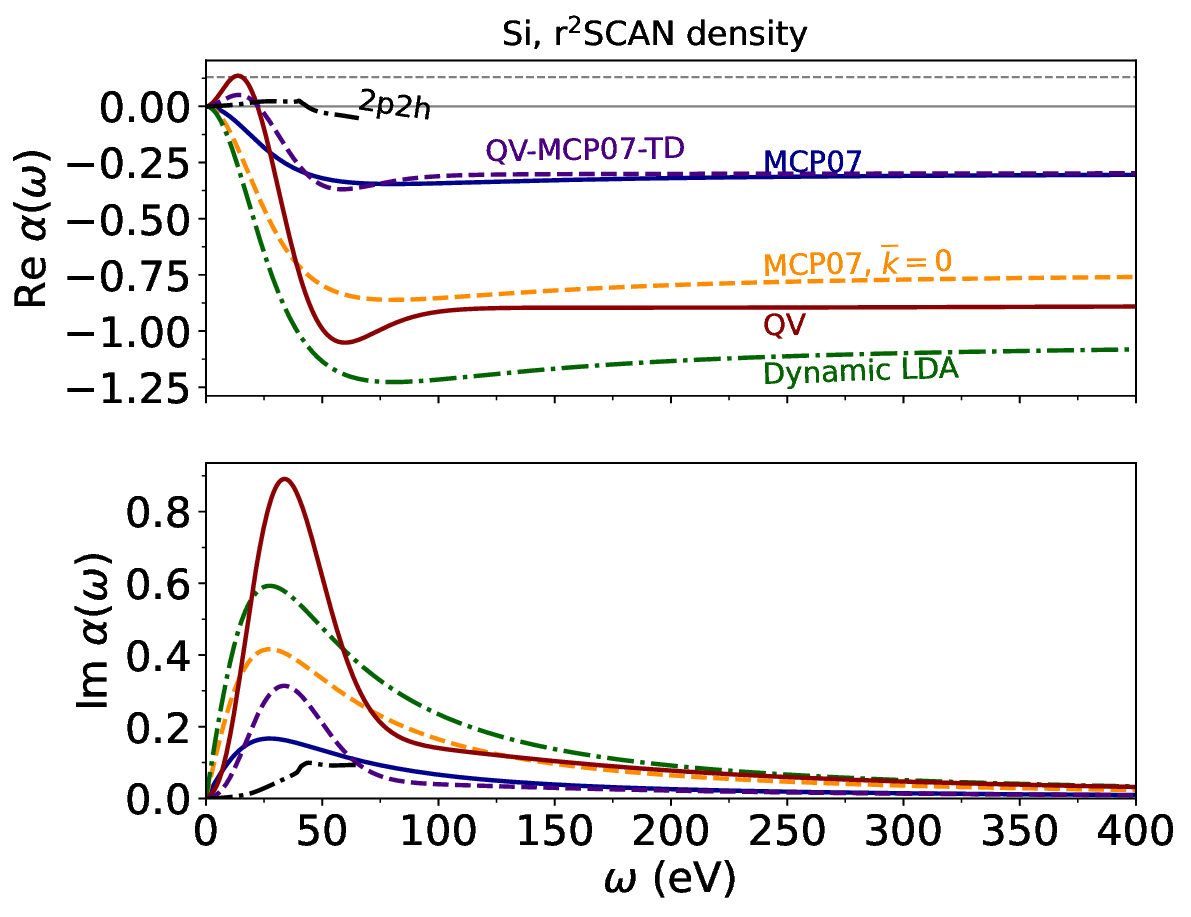}
  \caption{$\alpha(\omega)$ as a measure of ultranonlocality in cds Si crystal, calculated with the GKI dynamic LDA (green dot-dashed), MCP07 $\kb=0$ (orange dashed), MCP07 (blue solid), long-wavelength Qian and Vignale \cite{qian2002} (QV, red solid), QV-MCP07-TD (purple, dashed), and 2p2h \cite{panholzer2018} (black, dot-dashed) xc kernels. The horizontal gray dashed line is the static limit of the LRC ultranonlocality coefficient, $\alpha=0.13$ \cite{botti2005}. The QV kernel comes closest, at low frequencies, to approximating this value. \label{fig:si_alpha_bigrange}}
\end{figure}

\begin{figure}
  \includegraphics[width=\columnwidth]{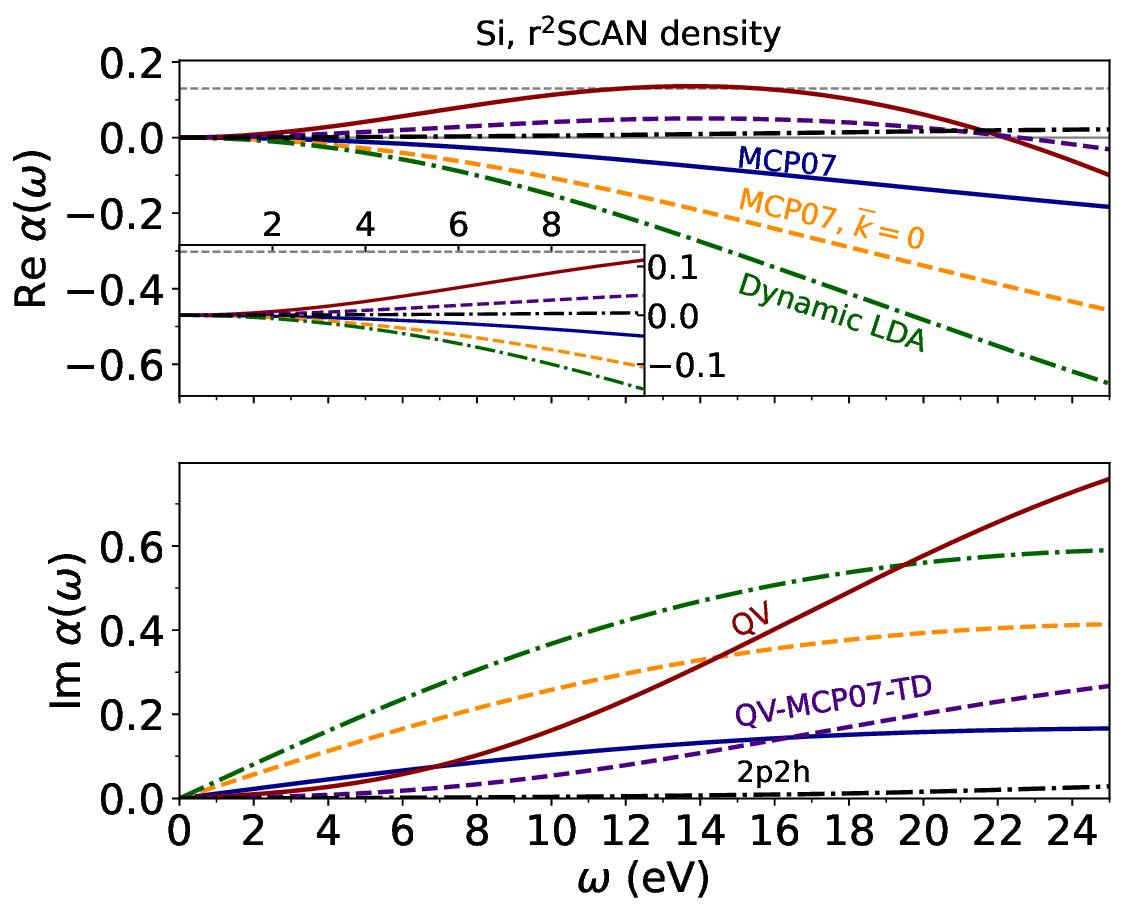}
  \caption{$\alpha(\omega)$ as a measure of ultranonlocality in cds Si crystal for the small omega regime only, calculated with the GKI dynamic LDA (green dot-dashed), MCP07 $\kb=0$ (orange dashed), MCP07 (blue solid), long-wavelength Qian and Vignale \cite{qian2002} (QV, red solid), QV-MCP07-TD (purple, dashed), and 2p2h \cite{panholzer2018} (black, dot-dashed) xc kernels. The horizontal gray dashed line is the static limit of the LRC ultranonlocality coefficient, $\alpha=0.13$ \cite{botti2005}. The inset shows the range $0 < \omega < 10$ eV. We emphasize that Ref. \cite{nazarov2009} used a different sign convention for $\alpha(\omega)$, $\fxc = e^2\alpha/q^2$. Therefore, Fig. 2 of Ref. \cite{nazarov2009}, which plots $\alpha(\omega)$ for Si, appears to have the sign of $\alpha(\omega)$ reversed.
 \label{fig:si_alpha}}
\end{figure}

\begin{figure}[t]
  \includegraphics[width=\columnwidth]{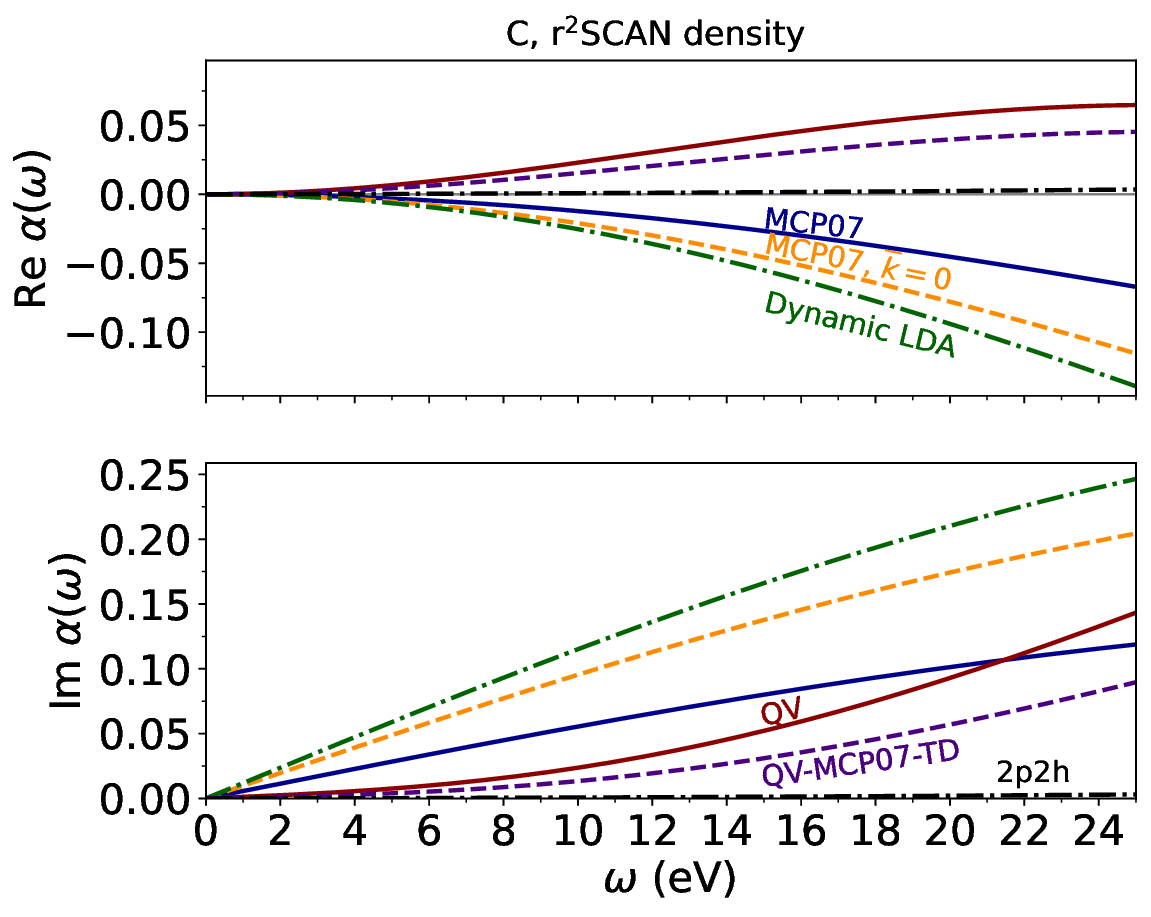}
  \caption{$\alpha(\omega)$ as a measure of ultranonlocality in cds C crystal, calculated with the GKI dynamic LDA (green dot-dashed), MCP07 $\kb=0$ (orange dashed), MCP07 (blue solid), long-wavelength Qian and Vignale \cite{qian2002} (QV, red solid), QV-MCP07-TD (purple, dashed), and 2p2h \cite{panholzer2018} (black, dot-dashed) xc kernels. The static limit of the LRC ultranonlocality coefficient, $\alpha=0.28$, \cite{botti2005} is beyond the scale of the vertical axis. As cds C has a much larger bandgap than cds Si, we begin to see the limited validity of applying Eq. (\ref{eq:alpha_sum}) to insulators. \label{fig:c_alpha}}
\end{figure}

Figures \ref{fig:si_alpha_bigrange}, \ref{fig:si_alpha}, and \ref{fig:c_alpha} present our calculations of $\alpha(\omega)$ for cubic diamond structure (cds) Si and C. In this case, we obtain reliable ground-state valence electron densities using a plane-wave basis set. These calculations employed r$^2$SCAN \cite{furness2020}, a computationally efficient and highly-accurate ground-state density functional, within the Vienna \textit{Ab initio} Simulation Package \cite{kresse1996,*kresse1994,*kresse1993,*kresse1996a}. Refer to Appendix A for definitions of the quantities $\overline{n}_0$ and $n(\bg)$ entering Eq. (\ref{eq:alpha_sum}). To ensure that a large number of wavevectors were used in the Hamiltonian, the calculation was performed on a $\Gamma$-centered $\bm{k}$-mesh of spacing 0.08 \AA{}$^{-1}$, with an energy cutoff of 800 eV. The tetrahedron integration method was used to obtain reliable total energies converged within $10^{-7}$ eV. The equilibrium volume was determined by fitting to the stabilized jellium equation of state \cite{staroverov2004}; r$^2$SCAN predicts an equilibrium cubic lattice parameter of 5.440 \AA{} for Si and 3.562 \AA{} for C. The real-space density was then Fourier transformed to yield $n(\bg)$ for reciprocal lattice vectors in the range $|\bg|^2/2 < 3200$ eV.

Our results for $\alpha(\omega)$ in semiconductors are presented in Figs. \ref{fig:si_alpha_bigrange} and \ref{fig:si_alpha} for cds Si, and Fig. \ref{fig:c_alpha} for cds C. At optical frequencies in the range 3--5 eV, the coefficient $\alpha(\omega)$ is positive, as expected, only when the frequency dependence of the xc kernel is taken to be that of Qian and Vignale. However, even the QV kernel considerably underestimates the expected values $\alpha = $0.2 \cite{botti2004} or 0.28 \cite{botti2005}. Note that the QV xc kernel uses the xc shear modulus $\mu\sxc$, which was tabulated only for $\rs = 1,2,3,4,5$ in Ref. \cite{qian2002}. To interpolate and extrapolate these values, we used the physically-motivated \cite{conti1999} form
\begin{equation}
    \frac{\mu\sxc(\rs)}{n} = \frac{a}{\rs} + (b-a)\frac{\rs}{\rs^2 + c}, \label{eq:xc_shear}
\end{equation}
with $a = 0.031152$, $b=0.011985$, and $c=2.267455$ fitted to the values reported in Ref. \cite{qian2002} (in atomic units). The Perdew-Wang (PW92) \cite{perdew1992} parametrization of the UEG correlation energy per electron $\varepsilon_{\mathrm{c}}$ was used as input to the static compressibility of the QV and QV-MCP07-TD kernels. For the GKI dynamic LDA and MCP07 with $\kb=0$ or $k$, we employed the Perdew-Zunger \cite{perdew1981} parametrization of $\varepsilon_{\mathrm{c}}$, consistent with Ref. \cite{ruzsinszky2020}.

For Si and C, we also present (when possible) the static limit of the ultranonlocality coefficient determined by the empirical long-range contribution (LRC) xc kernel of Ref. \cite{botti2005}:
\begin{equation}
    \fxc^{\text{LRC}}(q,\omega) = -\frac{\alpha + \beta \omega^2}{q^2}. \label{eq:fxc_lrc}
\end{equation}
Here, $\alpha$ and $\beta$ are material-dependent parameters that are fitted to spectroscopic data. For Si, $\alpha=0.13,\, \beta = 0.00635~\text{eV}^{-2}$; for C, $\alpha=0.28,\, \beta = 0.00135~\text{eV}^{-2}$. This gives a benchmark for the kernels presented here, and helps determine the validity of Eq. (\ref{eq:alpha_sum}) for insulators.

As an alternative to the LRC benchmark, we also consider the 2p2h kernel \cite{panholzer2018}. This kernel is determined directly from Fermi hypernetted chain calculations of the UEG, including two-particle, two-hole (2p2h) interactions. As these excitonic interactions are relevant for the optical regime, the 2p2h kernel may be the best point of reference for our work. See Appendix C for a discussion of the 2p2h kernel and its limitations.

The real part of the coefficient $\alpha(\omega)$, as defined in Eq. (\ref{eq:fxc_alpha}), is expected to vanish at $\omega = 0$ for metals and to be positive at $\omega = 0$ for semiconductors and insulators (e.g., +0.2 as in Ref. \cite{botti2004} for cds Si). Equation (\ref{eq:alpha_sum}), which is formally exact in the limit of weak inhomogeneity and is, therefore, suitable for metals, always yields $\alpha(\omega=0)=0$, as expected for metals. For $\omega$ in the visible range, we expect $\re  \alpha(\omega)$ to be positive (and small for metals). The computed sign of $\re \alpha(\omega)$ turns out, however, to be positive in the visible range only when the QV or 2p2h frequency dependence is used. The MCP07 xc kernel of Eq. (\ref{eq:mcp07_fxc}) represents an extension, for finite wavevectors, of the GKI dynamic xc kernel, which differs considerably at small $\omega$ from its QV counterpart: they satisfy distinct $\omega \to 0$ limits, and thus have different Taylor expansions near $\omega = 0$.
These important differences should be at the origin of the different behavior of $\alpha(\omega)$, particularly at small $\omega$, depending on whether the QV dynamic LDA xc kernel is used.

While $\re \alpha(\omega)$ is predicted to be negative (and not positive as expected) at small frequencies for the three GKI-based kernels under study, it does become (plausibly for metals) smaller in magnitude as we make the wavevector dependence of this kernel increasingly sophisticated (from dynamic LDA to MCP07 $\kb=0$ to MCP07 $\kb = k$). Our work reveals the remarkable sensitivity of Eq. (\ref{eq:alpha_sum}) to the wavevector and frequency dependence of the UEG xc kernel.

The long-range part of the xc kernel is expected to be highly non-monotonic in its frequency dependence \cite{delsole2003}. This was demonstrated in Ref. \cite{delsole2003} for cds Si and cds C using the response function computed from both the Bethe-Salpeter equation and ground-state LDA eigenstates. While all kernels presented here show a nontrivial frequency dependence, none
of them demonstrate the anticipated oscillatory behavior of $\alpha(\omega)$.

We do not advocate using MCP07 in its current form for general optical applications; however we recommend further testing of the QV-MCP07-TD kernel for the study of optical properties. This novel kernel demonstrates that introducing new constraints beyond the GKI-based frequency dependence could improve MCP07's predictions of optical properties. Our aim here is to highlight the relevance of a proper description of both the frequency and wavevector dependence of $\fxc$ in the optical limit. Future work could study the application of the QV-MCP07-TD kernel to describe the optical properties of a broader range of semiconductors.

\begin{figure}
  \includegraphics[width=\columnwidth]{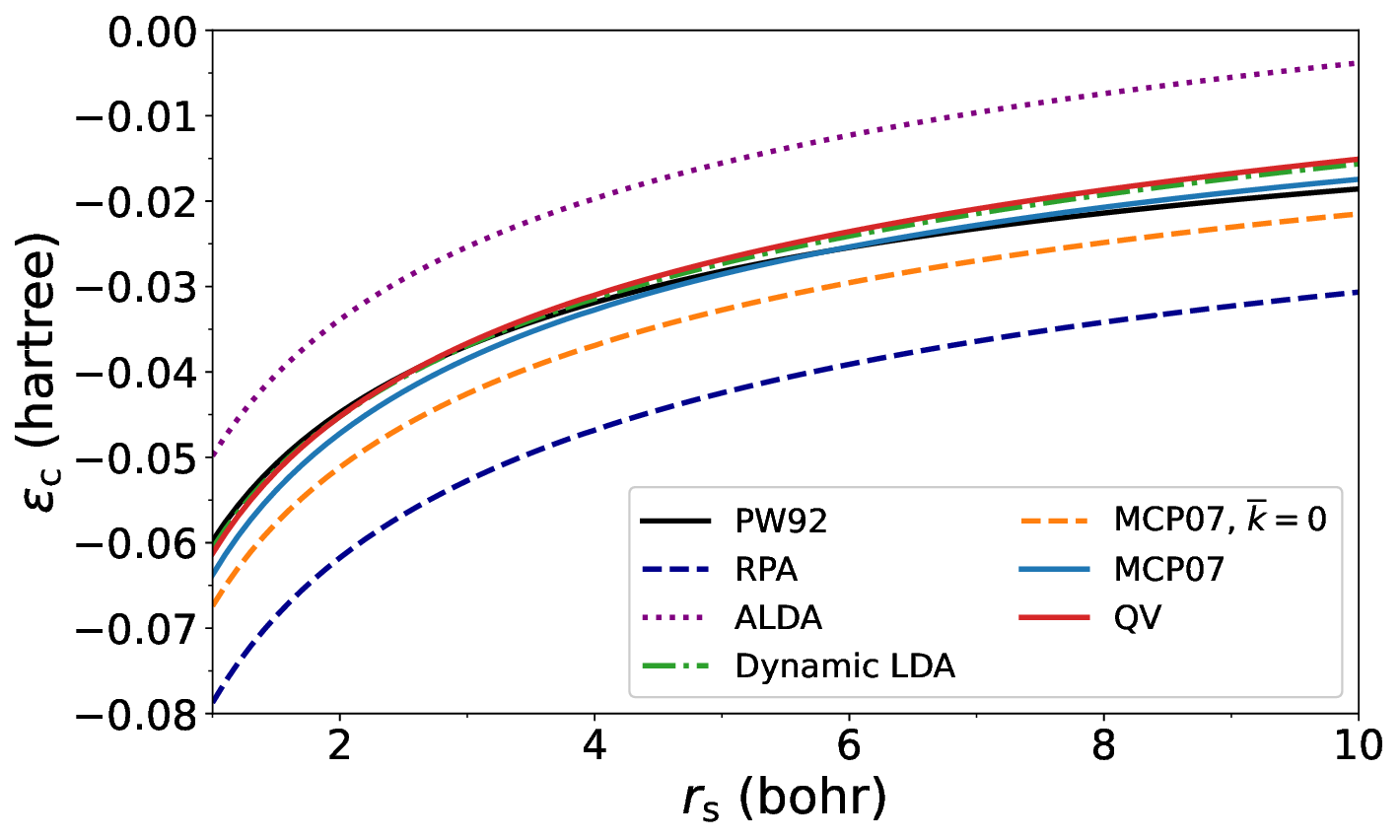}
  \caption{ The correlation energy per electron of the UEG as a function of the density parameter $\rs$. Besides the PW92 ground-state LDA (black solid), the figure displays the ALDA (purple dotted), RPA (dark blue dashed), MCP07 (light blue solid), MCP07 $\kb=0$ (orange dashed), the GKI dynamic LDA (dark green dash-dotted), and the QV dynamic LDA (red solid) xc kernels. See Fig. \ref{fig:ueg_eps_c_errs} and Table \ref{tab:ueg_eps_c_errs} in Appendix B for a discussion of the relative errors made by these and other kernels.
  \label{fig:ueg_eps_c}}
\end{figure}

To complete the analysis of the coefficient $\alpha(\omega)$, we have performed a calculation of the ground-state correlation energy of the UEG for electron-density parameters in the range from $\rs = 1$ to $\rs = 10$. The PW92 approximation \cite{perdew1992} (black solid line in Fig. \ref{fig:ueg_eps_c}) is regarded as exact. In Fig. \ref{fig:ueg_eps_c}, we plot the full MCP07 correlation energy (blue solid line in Fig. \ref{fig:ueg_eps_c}) together with the result of using various approximations to $\fxc$: the fully non-local dynamic MCP07 with $\kb=0$, the GKI dynamic LDA, the QV dynamic LDA, the ALDA, and the RPA. The method for obtaining correlation energies through the fluctuation dissipation theorem \cite{furche2005}, using the Cauchy Integral Formula for the frequency integral, is described in Ref. \cite{ruzsinszky2020}. Here we see that while the ALDA is known to overestimate the correlation energy significantly for all values of $\rs$, introducing frequency dependence (still within the LDA, $q=0$ in $\fxc$) improves the correlation energy considerably. The fully non-local and dynamic MCP07 xc kernel yields excellent correlation energies, particularly with increasing $\rs$. When the frequency dependence of the MCP07 xc kernel is undamped by taking $\kb=0$, the correlation energy is clearly worsened, but not to the same extent as in the case of $\alpha(\omega)$.

Note, however, the role of the wavevector in determining accurate correlation energies. Both the QV kernel and dynamic LDA, which are wavevector-independent, tend to over-correct the RPA. However, the MCP07 $\kb=0$ kernel, which adds a naive wavevector dependence to the dynamic LDA, gives less reliable estimates of the UEG correlation energy. Using nonzero $\kb$ and correctly interpolating between known limits, as in MCP07, improves upon the dynamic LDA. It is easy to show that the analytic continuation of $\fxc(\omega)$ to imaginary frequency $iu$ (with $\omega$ and $u$ purely real), which is needed for the calculation of correlation energies, is
\begin{align}
\re \fxc(iu) &= f_{\infty} + \frac{1}{2\pi}\int_{-\infty}^{\infty} [\omega'^2 + u^2]^{-1} \{[\re \fxc(\omega')\nonumber \\
  & -f_{\infty}] u + \im \fxc(\omega')\omega'\} d\omega'  \label{eq:re_fxc_im_freq} \\
  \im \fxc(iu) &= \frac{1}{2\pi}\int_{-\infty}^{\infty} [\omega'^2 + u^2]^{-1} \{-[\re \fxc(\omega')-f_{\infty}] \omega' \nonumber \\
  &+ \im \fxc(\omega')u\} d\omega', \label{eq:im_fxc_im_freq}
\end{align}
with $f_{\infty} \equiv \lim_{\omega\to\infty} \fxc(\omega)$. The GKI and QV dynamic LDAs make $\im \fxc(\omega)$ a purely odd function of real-valued frequency $\omega$, and thus their corresponding $\re \fxc(\omega)$ are even functions of $\omega$. From Eq. (\ref{eq:im_fxc_im_freq}), we see that the integrand of $\im \fxc(iu)$ is odd in
$\omega'$, and thus $\fxc(iu)$ is purely real.

\section{Conclusions}

We have used the dynamic MCP07 kernel to study the roles of the wavevector and frequency dependence of $\fxc$ in the optical limit. Both ingredients have a significant impact in the visible region of light.

The strong wavevector dependence of $\alpha(\omega)$ corroborates the presence of local-field effects that have been discussed earlier. Less is known about the role of the frequency dependence in the optical limit.

To describe local-field effects, a kernel with the correct ultranonlocal limit is needed. Although there are a few efforts listed in the literature, no non-empirical xc kernels within TDDFT yield universal applicability for the optical absorption spectra of semiconductors and insulators. The TDCDFT framework allows the construction of, in the optical limit, a frequency-dependent xc kernel of a weakly inhomogeneous electron gas that exhibits ultranonlocality, as anticipated in Ref. \cite{nazarov2009}. In this work, we rely on this formalism to evaluate the coefficient $\alpha(\omega)$ as a measure of ultranonlocality in the optical limit.

We have included three types of xc kernels in our analysis.  As potentially better references, we have also included the dynamic LDA of Qian and Vignale \cite{qian2002} and the so-called 2p2h kernel \cite{panholzer2018}. Although there is no reference about what the correct ultranonlocality coefficient in the optical limit is, the completely local dynamic LDA sets a negative extreme with the GKI frequency model. Our work demonstrates the relevance of exact constraints. An approximate kernel can improve beyond the ALDA [which makes $\alpha(\omega)$=0] by enforcing known limiting behaviors of the exact $\fxc(q,\omega)$, as the GKI, QV, and MCP07 kernels are constructed to do. The MCP07 kernel is an interpolation between the static and frequency-only limits. The impact of wavevector and frequency have been independently investigated using the MCP07 model. Turning off the wavevector and frequency-dependence reduces all UEG-based xc kernels to the ALDA. Turning off the damping factor $\kb$ in MCP07 results in a less-controlled frequency dependence, but the resultant kernel retains the correct wavevector dependence of MCP07 in the static limit. The resulting $\alpha(\omega)$ is significantly reduced compared to that of the ALDA, indicating a reduced degree of ultranonlocality. The full MCP07 kernel damps the frequency dependence of the GKI dynamic LDA, and reduces $\alpha(\omega)$ further.

We have tested all these kernels for fcc Al, bcc Na, cds Si, and cds C. These are metallic and semiconducting systems with small ultranonlocality for metals, and larger ultranonocality for semiconductors. For the metallic systems, perturbation theory applies. We have obtained novel and informative results as an estimation of the ultranonlocality for metals. Our analysis confirms that even metals can have ultranonlocality, although to a much lesser extent than semiconductors.

Neither the dynamic LDA nor MCP07 exhibit the estimated level and sign of ultranonlocality for Si and especially for diamond C. Our results are qualitative and indicate a strong sensitivity of $\alpha(\omega)$ in the optical limit to the frequency and wavevector-dependence of the uniform gas kernel. This sensitivity was out of the reach in Ref. \cite{nazarov2009}, as only the QV dynamic LDA kernel was used, in contrast with the various non-local kernels we have considered here.  Furthermore, the non-local MCP07 and QV-MCP07-TD kernels considered here appear to be a reasonable basis for further improvements, especially when compared with the dynamic LDA. Our work clearly indicates  the limitations of the GKI frequency model. Instead, the hybrid QV-MCP07-TD kernel introduced here unifies both wavevector and frequency constraints, and shows a good promise within the limitations of the perturbation theory applied throughout this work.

Our current work could (i) guide further modifications in the frequency-dependent MCP07 kernel for optical spectroscopy to build in more exact constraints on the frequency dependence and (ii) guide more efforts to extend the TDCDFT scheme to obtain $\alpha(\omega)$ and the corresponding $\lim_{q\to0}\fxc(\bm{q},\bm{q};\omega)$ for non-uniform systems.

\begin{acknowledgments}

    The authors are grateful for fruitful discussions with Professor John P. Perdew.

  AR and NKN acknowledge support from the U.S. National Science Foundation under Grant No. DMR-1553022.  ADK acknowledges support from the U.S. Department of Energy, Basic Energy Sciences, through the Energy Frontier Research Center for Complex Materials from First Principles Grant DE-SC0012575; and Temple University.

\end{acknowledgments}

\section*{Author Contributions}

N.K.N. and A.D.K. performed the calculations and revised the paper. J.M.P. and A.R. designed the research and revised the paper. A.R. wrote the paper.

\section*{Data and code availability}

The Python code used to generate Figs. \ref{fig:al_alpha}--\ref{fig:c_alpha} from VASP outputs is made freely available (without access restrictions) at \url{https://github.com/esoteric-ephemera/ultranonlocal_semiconductors}, under the ``code'' directory. The raw data from VASP is also included there, under the ``Si'' and ``C'' directories. The processed data is located in the ``code/data\_files'' subdirectory.


%

\appendix

\section{Definitions of density variables}

Consider a density $n(\br)$ that varies weakly about an average uniform density $\overline{n}_0$. As is done in all standard electronic structure codes, we sample the (self-consistent) density in real space at $N_r$ points $\bm{R}$. To obtain the Fourier components $n(\bg)$, we take the discrete Fourier transform
\begin{equation}
    n(\bg) = \frac{1}{N_R}\sum_{\bm{R}} n(\bm{R}) e^{-i \bg\cdot \bm{R}},
\end{equation}
leveraging the fast Fourier transform for a suitable choice of $\bm{R}$,
\begin{align}
  \bm{R} &= \frac{n_1}{N_1} \bm{a}_1 + \frac{n_2}{N_2} \bm{a}_2 + \frac{n_3}{N_3} \bm{a}_3 \\
  \bg &= m_1 \bm{b}_1 + m_2 \bm{b}_2 + m_3 \bm{b}_3 \\
  \bg \cdot \bm{R} &= 2\pi\left(\frac{n_1 m_1}{N_1} + \frac{n_2 m_2}{N_2} + \frac{n_3 m_3}{N_3} \right),
\end{align}
where $n_i,\, m_i = 0,1,2,...,N_i-1$, such that $N_1 N_2 N_3 = N_R$. $\bm{a}_i$ are the direct lattice vectors, and $\bm{b}_j$ are the reciprocal lattice vectors such that $\bm{a}_i\cdot \bm{b}_j = 2\pi \delta_{ij}$. This convention is adopted by VASP; however, any standard plane-wave code will make similar choices for the discrete Fourier transform phase and normalization conventions, and will make similar choices for $\bm{R}$.

Real solids are typically not weakly varying; however, the valence densities of Si, and less so C, are approximately weakly varying about their average density, which we \emph{define} as
\begin{equation}
    \overline{n}_0 \equiv n(\bg=\bm{0}).
\end{equation}
In pseudopotential codes like VASP, only valence electrons are considered, and the core electrons are replaced by a non-local effective potential inside a core radius. Therefore, the variables $n(\bm{R})$ and $n(\bg)$ represent the \emph{valence} electron density and the Fourier transform of the valence electron density, respectively.

Virtually, all xc kernels based on the UEG paradigm require a real-density input; therefore, in Eq. (\ref{eq:alpha_sum}) we evaluate
\begin{equation}
    \fxc(q,\omega) \equiv \fxc(\overline{n}_0,q,\omega).
\end{equation}
Let $\overline{\rs}^3 = 3/(4\pi \overline{n}_0)$. In our self-consistent calculations, we found $\overline{\rs} \approx 2.009$ bohr for cds Si (which is within a reasonable metallic range, $2 \lesssim \overline{\rs} \lesssim 5$ bohr), and $\overline{\rs} \approx 1.315$ bohr for cds C (which is outside the normal metallic range).

\section{A more detailed discussion of the correlation energies}

In this Appendix, we compare a variety of xc kernels in predicting jellium correlation energies (per electron) for the physically relevant range of electron densities
$1 \leq \rs \leq 10$. We take PW92 \cite{perdew1992} to be essentially exact.

The correlation energy per electron, $\varepsilon_{\mathrm{c}}$ can be computed from the adiabatic-connection fluctuation-dissipation theorem: \cite{langreth1977}
\begin{equation}
  \varepsilon_{\mathrm{c}} = \frac{1}{2} \int \frac{d^3 q}{(2\pi)^3} \int_0^1 \frac{d\lambda}{\lambda} \int_0^{\infty} d\omega \frac{4\pi \lambda }{q^2}[ S_{\lambda}(\bq,\omega) - S_0(\bq,\omega) ],
\end{equation}
where $f_{\mathrm{xc},\lambda}(\bq,\omega,\rs) = \lambda^{-1}\fxc(\lambda^{-1} \bq, \lambda^{-2} \omega, \lambda \rs)$ \cite{lein2000} and the spectral function $S_{\lambda}$ at coupling constant $\lambda$ is given by
\begin{align}
    \chi_{\lambda}(\bq,\omega) &= \frac{\chi_0(\bq,\omega)}{1 - [4\pi\lambda /q^2 + f_{\mathrm{xc},\lambda}(\bq,\omega,\rs)] \chi_0(\bq,\omega)}\\
    S_{\lambda}(\bq,\omega) &= -\frac{1}{\pi n}\im \chi_{\lambda}(\bq,\omega).
\end{align}
In the random phase approximation (RPA), $\fxc$ is taken to be zero. In this approximation, correlation energies are found to be too negative \cite{lein2000}. The use of appropriate approximations for $\fxc$ should build upon the RPA and improve its prediction of correlation energies.

This section also presents a few new kernels. In addition to those presented previously, we also consider the static limit of the MCP07 kernel, and the dynamic kernel formed by replacing, within MCP07, the GKI dynamic LDA with the QV dynamic LDA. This combination can be done in two ways, as the TDDFT and TDCDFT static and $q\to 0$ limits are incompatible \cite{conti1999}:
\begin{align}
    \lim_{q\to 0} \left[\lim_{\omega \to 0} \fxc^\text{HL}(q,\omega)\right] &= \fxc^{\text{GKI}}(0) = \fxc^{\text{ALDA}},\qquad \text{TDDFT} \\
    \lim_{\omega\to 0}\left[\lim_{q \to 0} \fxc^\text{HL}(q,\omega)\right] &= \fxc^{\text{QV}}(0) = \fxc^{\text{ALDA}} + \mu\sxc,\quad \text{TDCDFT}
\end{align}
where $\fxc^\text{HL}(q,\omega)$ is the longitudinal xc kernel of a homogeneous electron gas within TDCDFT and $\mu\sxc$ is the xc shear modulus [see Eq. (\ref{eq:xc_shear})]. For consistency, the two new QV-MCP07 kernels use the PW92 parametrization of the correlation energy, and have the form
\begin{equation}
    \fxc(q,\omega) = \left\{ 1 + e^{-\kb q^2}\left[\frac{\fxc^{\text{QV}}(\omega)}{\fxc(0,0)} - 1\right] \right\}\fxc^{\text{MCP07}}(q,0).
\end{equation}
To obtain the QV-MCP07-TD kernel, we take $\fxc(0,0)=\fxc^{\text{ALDA}}$ in the MCP07 static kernel and the QV kernel, by setting $\mu\sxc=0$. To obtain the QV-MCP07-TDC kernel, we take $\fxc(0,0)=\fxc^{\text{ALDA}} + \mu\sxc(\rs)$, which modifies $\kb$ and $\fxc^{\text{MCP07}}(q,0)$, consistent with the construction principles of Ref. \cite{ruzsinszky2020}.

\begin{figure*}
    \centering
    \includegraphics[width=\textwidth]{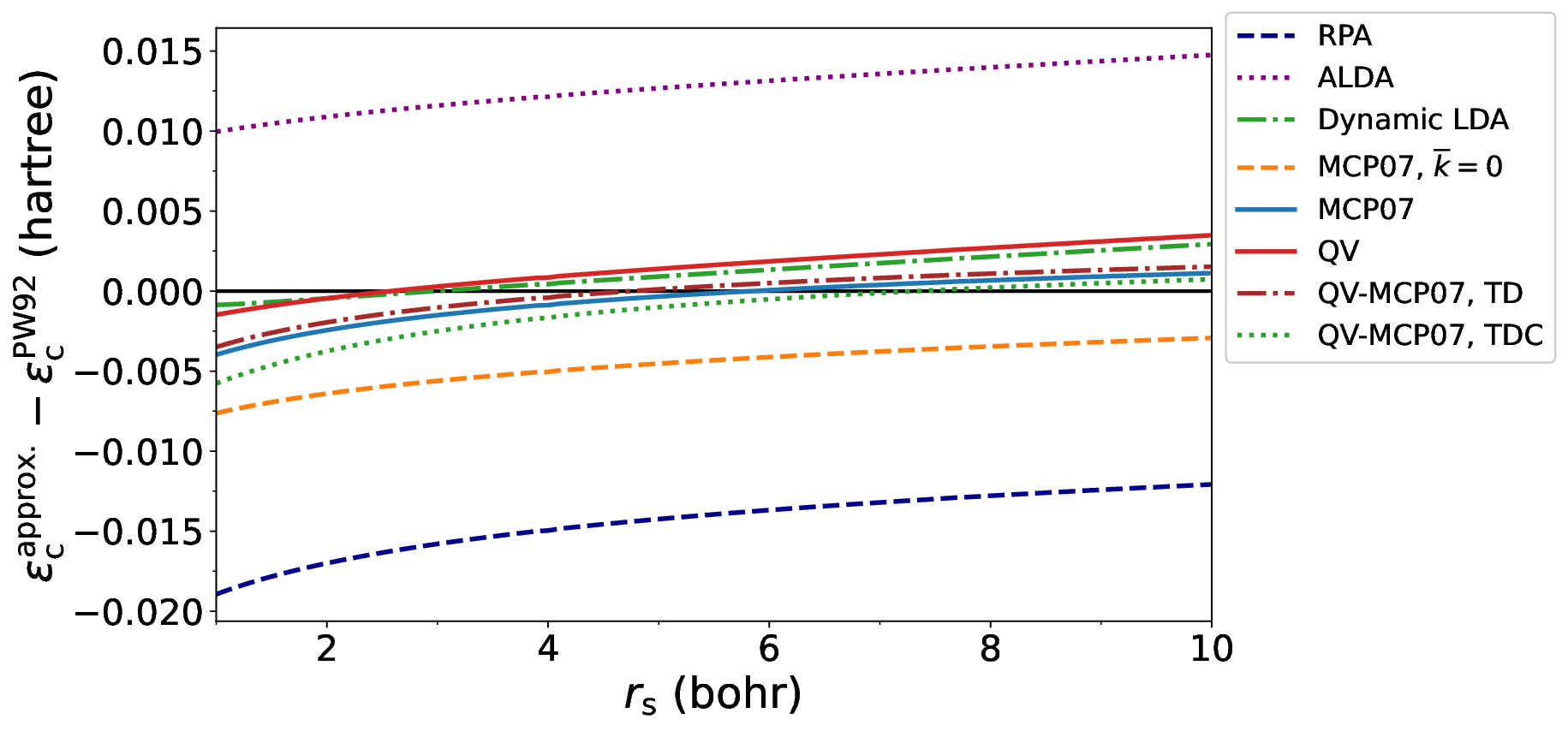}
    \caption{Comparison of the relative errors $\varepsilon^{\text{approx.}}_{\text{c}}-\varepsilon^{\text{PW92}}_{\text{c}}$ for a variety of functionals. Error statistics are given in Table \ref{tab:ueg_eps_c_errs}.}
    \label{fig:ueg_eps_c_errs}
\end{figure*}

Figure \ref{fig:ueg_eps_c_errs} plots the relative errors in the correlation energies per electron as a function of $1 \leq \rs \leq 10$. Table \ref{tab:ueg_eps_c_errs} presents the average errors and standard deviations in the UEG correlation energy for the same range of $\rs$. It can be seen that the most accurate kernels in this range are the MCP07 and QV-MCP07-TD kernels, whose mean absolute errors are comparable, differing only by about $4 \times 10^{-6}$ hartree.

\begin{table*}[t]
\begin{ruledtabular}
    \centering
    \begin{tabular}{rp{2.5cm}p{4cm}p{3.2cm}}
      Kernel & Mean error & Mean absolute error & Standard deviation \\
       & $\times 10^{-2}$ hartree & $\times 10^{-2}$ hartree &  $\times 10^{-3}$ hartree \\ \hline
      RPA & -1.4387 & 1.4387 & 1.8092 \\
      ALDA & 1.2750 & 1.2750 & 1.3251 \\
      Dynamic LDA & 0.1082 & 0.1289 & 1.1242 \\
      MCP07 static & 0.0836 & 0.1277 & 1.1603 \\
      MCP07, $\overline{k}=0$ & -0.4592 & 0.4592 & 1.2591 \\
      MCP07 & -0.0496 & 0.1077 & 1.3382 \\
      QV & 0.1455 & 0.1705 & 1.3580 \\
      QV-MCP07, TD & -0.0047 & 0.1074 & 1.3104 \\
      QV-MCP07, TDC & -0.1251 & 0.1491 & 1.7068
    \end{tabular}
    \caption{Comparison of approximate exchange-correlation kernels in predicting jellium correlation energies (per electron), for 91 values of $\rs$ in the range $1 \leq \rs \leq 10$. PW92 \cite{perdew1992} is taken to be the reference energy.}
    \label{tab:ueg_eps_c_errs}
\end{ruledtabular}
\end{table*}

\section{Technical aspects of the 2p2h kernel calculation}

The 2p2h kernel, while capturing a broad range of many-electron physics, is tabulated only for a limited range of $\rs$, $q$, and $\omega$. Furthermore, no analytic expression has been determined to interpolate it. Thus, we are forced to make two approximations to use the 2p2h kernel in a practical computation.

To interpolate the kernel, we use a multivariate linear interpolation. For a one-dimensional function $F(x)$, this amounts to
\begin{align}
    F(x) &\approx F(x_i)\frac{x_{i+1}-x}{x_{i+1} - x_{i}} + F(x_{i+1})\frac{x - x_i}{x_{i+1} - x_{i}}, \\
    x_i &\leq x < x_{i+1},
\end{align}
and the tabulated values of $x_i$ are sorted by increasing values. In three dimensions, we perform a simple composition of one-dimensional linear interpolations. More sophisticated multidimensional interpolation schemes, like tri-cubic spline, assume the function to be interpolated is smooth in some sense: e.g., a cubic spline assumes continuity up to the second derivatives. We cannot make such an assumption about $\fxc(\rs,q,\omega)$.

\begin{ruledtabular}
    \begin{table}[ht]
        \centering
        \begin{tabular}{ccrr}
            Solid (structure) & $\rs$ (bohr) & $q^2_{\text{cut}}/2$ (eV) & $\omega_{\text{cut}}$ (eV) \\ \hline
            C  (cds) &  1.32  &  231.68  &  124.33  \\
            Al  (fcc) &  2.07  &  93.56  &  62.99  \\
            Si  (cds) &  2.01  &  99.35  &  65.89  \\
            Na (bcc)  &  3.93  &  25.96  &  24.08  \\
        \end{tabular}
        \caption{Summary of cutoff energies used in the computation of $\alpha(\omega)$ for the 2p2h kernel only. For all other kernels, a cutoff of $|\bg|^2/2 < 800$ eV was used. Note that $q_{\text{cut}} = 8 \kf$, and $\omega_{\text{cut}} = 3.98\omega_p(0)$.}
        \label{tab:2p2h_cutoffs}
    \end{table}
\end{ruledtabular}

Second, the 2p2h kernel is tabulated only up to $q = 8 \kf$ and $\omega = 3.98 \omega_p(0)$, where $\omega_p(0) = (3/\rs^3)^{1/2}$ is the semiclassical plasmon frequency. These values are too small for  the 2p2h kernel to attain its zero-separation and infinite frequency limits. While the limiting values $\fxc(q= 0,\omega \to \infty)$ \cite{iwamoto1987} and $\fxc(q\to \infty,\omega = 0)$ \cite{corradini1998} are known, we cannot, in general, extrapolate to $q\to \infty$ with $\omega > 0$, or vice versa. Thus, we are forced to cut off the $\bm{G}$ sum of Eq. (\ref{eq:alpha_sum}) for $|\bg| > 8 \kf$, and restrict $\alpha(\omega)$ to $\omega \leq 3.98 \omega_p(0)$. The numeric values of the cutoffs are given in Table \ref{tab:2p2h_cutoffs}. Note that, for all solids considered here, the shortest reciprocal lattice vector is longer than the smallest wavevector ($0.1\kf$) for which the 2p2h kernel is tabulated.

\end{document}